\documentclass{aa}
\usepackage{astron}
\usepackage{times,graphicx}
\usepackage{supertabular}
\usepackage{curves,epic,eepic}
\usepackage{float}

\begin{document}

\thesaurus{03.13.5; 
           13.07.1;} 
         \title{Early detection of the optical counterpart to GRB
           980329~\thanks{Based on observations carried out at the 1.0-m
           Jacobus Kapteyn Telescope operated by the Royal Greenwich
           Observatory in the Spanish Observatorio del Roque de los
           Muchachos of the Instituto de Astrof\'{\i}sica de
           Canarias.}$^{,}$\thanks{Based on observations carried out at the
           1.5-m Telescope of the Observatorio Astron\'omico Nacional in
           Calar Alto.}  $^{,}$\thanks{Based on observations carried out at
           the 1.0-m Optical Ground Station telescope of the European Space
           Agency in the Spanish Observatorio del Teide of the Instituto de
           Astrof\'{\i}sica de Canarias.}}
   \author{J. Gorosabel
          \inst{1}
   \and A. J. Castro-Tirado
          \inst{1,2}
   \and A. Pedrosa
          \inst{3}
   \and M. R. Zapatero-Osorio          
          \inst{4}
   \and A. J. L. Fernades   
          \inst{3,5}
   \and M. Feroci
        \inst{6}
   \and E. Costa
        \inst{6}
   \and F. Frontera
        \inst{7,8}
}
   \offprints{J. Gorosabel (jgu@laeff.esa.es)}

\institute{Laboratorio de Astrof\'{\i}sica Espacial y F\'{\i}sica
     Fundamental (LAEFF-INTA), P.O. Box 50727, E-28080 Madrid, Spain.
\and Instituto de Astrof\'{\i}sica de Andaluc\'{\i}a (IAA-CSIC), P.O. Box
     03004, E-18080 Granada, Spain.
\and Centro de Astrofisica da Universidade do Porto, Rua das Estrelas s/n 15, 
4150 Porto, Portugal.
\and Instituto de Astrof\'{\i}sica de Canarias, La Laguna, Tenerife, Spain.
\and Intituto Superior da Maia, 4470 Maia, Portugal.
\and Istituto di Astrofisica Spaziale, CNR, Frascati, Italy.
\and Istituto Tecnologie e Studio Radiazioni Extraterrestri, CNR, Via P. Gobetti, 
I-40129 Bologna, Italy.
\and Dipartamento di Fisica, Universit\`a di Ferrara, I-44100 Ferrara, Italy.}

   \date{Received date; accepted date}
   
   \titlerunning{Detection of the GRB 980329 optical counterpart}

   \authorrunning{Gorosabel et al.}

   \maketitle

   \begin{abstract} We report optical observations  of the GRB 980329 error
     box which represent the second earliest detection of the GRB 980329
     optical transient.  We determine R = 23.7 $\pm$ 0.5 mag on 29.89
     March, which is consistent with R = 23.6 $\pm$ 0.2 mag as reported by
     Palazzi et al. (1998) on 29.99 March.  Based on extrapolations of the
     light curve we claim that the R-band magnitude of the GRB 980329 host
     galaxy should lie in the range 26.8 mag $<$ R $<$ 29 mag.  We also
     discuss the similarities with GRB 970111.  
   \end{abstract}

%

\section{Introduction}

The Gamma-Ray Burst Monitor (GBM) on board the X-ray satellite {\it BSAX}
was triggered by a gamma-ray burst (GRB) on 29 March 1997, UT 3 h 44 m 30 s
and localised by the Wide Field Camera (WFC) on {\it BSAX} at
$\alpha_{2000}=7^{h}02^{m}41^{s},
\delta_{2000}=38^{\circ}50^{\prime}42^{\prime \prime}$ (uncertainty
$3^{\prime}$ radius).  GRB 980329 was very intense and reached a peak
intensity of 6 Crab in the 2-26 keV range, being the second brightest burst
localised so far with the WFCs (Frontera et al. 1998).  The event was also
detected by the burst and transient source experiment (BATSE) on board the
{\it Compton Gamma-Ray Observatory}, which provided a position consistent
with the one given by the WFC (Briggs et al. 1998).  A follow-up
observation was initiated within 7 hr with the {\it BSAX} narrow-field
instruments (NFI).  The observation revealed a previously unknown X-ray
source, 1SAX J0702.6+3850, that faded by a factor of $\sim$ 3 over 14 hr of
observation (in\' \rm t Zand et al. 1998).  Radio observations performed
with the VLA at 8.4 GHz resulted in the detection of a variable radio
counterpart that peaked $\sim$ 3 days after the burst (Taylor et al.
1998a,b).  This detection allowed observers to re-examine images taken on
the first night after the gamma event.  Thus, observations at the position
of the radio counterpart revealed a fading source in the I- (Klose 1998),
R- (Palazzi et al. 1998, Pedersen et al. 1998), and K-bands (Larkin et al.
1998a,b; Metzger 1998) as well as in submillimetre wavelengths (Smith \&
Tilanus 1998a,b).

The data provided by the Interplanetary Network (IPN) allowed to derive an
annulus that intersected the WFC error and reduced its area (Hurley et al.
1998). The new error box included the 1SAX J0702.6+3850 X-ray source, the
variable radio source and the optical transient (OT).  Deep optical
observations revealed a faint galaxy (R $= 25.7 \pm 0.3$ mag) coincident to
the optical fading source (Djorgovski et al. 1998), that was pointed out as
the GRB 980329 host galaxy. Here we present the result of the optical
observations performed at several observatories.

\section{Observations and data analysis}

Optical images in the R and B bands were acquired with three telescopes,
namely the 1.5-m OAN at the German-Spanish Observatorio de Calar Alto, the
1.0-m JKT at Observatorio del Roque de los Muchachos and the 1.0-m OGS
telescope at Observatorio de Iza\~na.

\begin{figure*}[t]
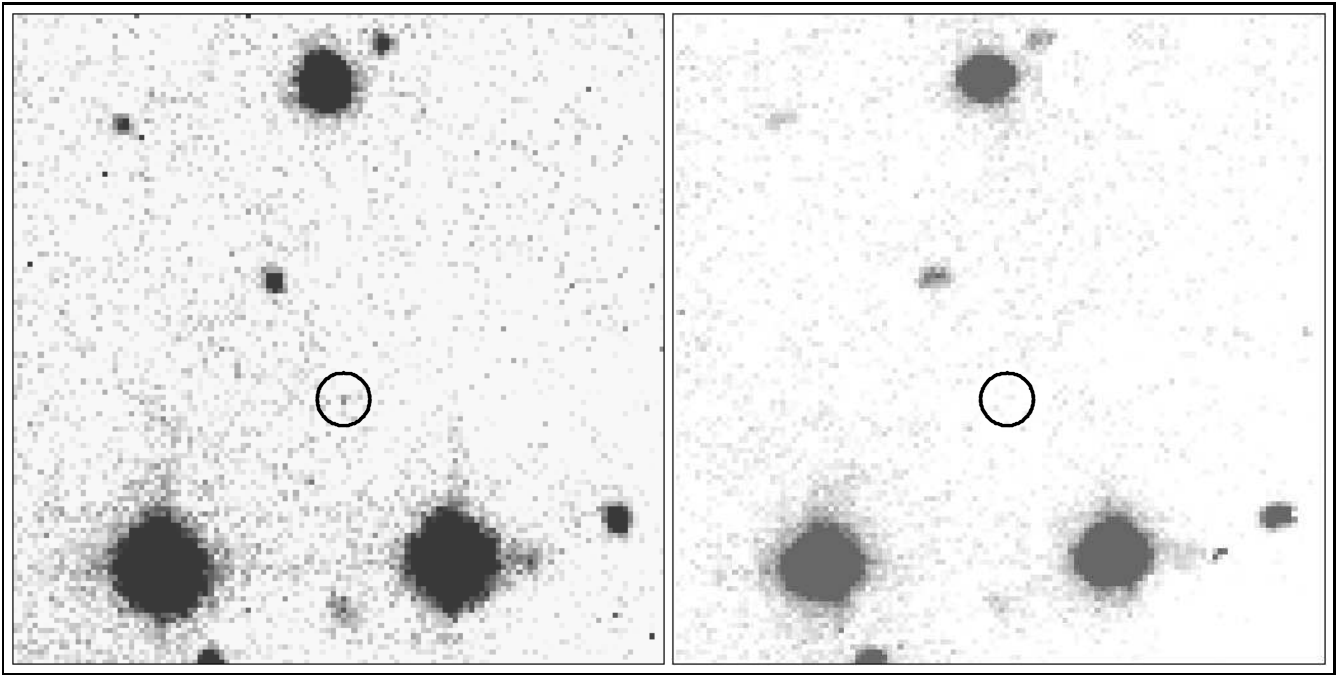

  \centering
\fbox{\resizebox{8.7cm}{8.7cm}{\rotatebox[origin=c]{180}{\includegraphics{Bb263.f1}}}
\resizebox{8.7cm}{8.7cm}{\rotatebox[origin=c]{180}{\includegraphics{Bb263.f2}}}}
    \linethickness{0.3mm}
  \put(-375,101){\bigcircle{20}}
  \put(-124,101){\bigcircle{20}}
\caption[R-band image of the counterpart to GRB 980329] {\label{980329fig2}
  The image shows the two R-band images of GRB 980329 taken with the JKT
  telescope; on 29.89 March (left) and the co-added image on 31.94 March.
  (right).  The long integration times of 2700 s and 3600 s allowed to
  reach 3$\sigma$ limiting magnitudes of R = 22.8 mag and 23.0 mag
  respectively. The faint object present at the radio counterpart position
  (circle), is not detected in the image taken on 31.94 March. The FOV is
  $38^{\prime \prime} \times 38^{\prime\prime}$, with north is to the top
  and east to the left.}
\end{figure*}

The 1.5-m OAN telescope is located at Calar Alto and its CCD provides a
field of view (FOV) of $6^{\prime}.9 \times 6^{\prime}.9$.  The 1-m JKT
telescope, placed at La Palma, gives a FOV of $5^{\prime}.6 \times
5^{\prime}.6$. The European Space Agency OGS is a 1-m telescope at the
Observatorio del Teide. Its camera provides a $5^{\prime}.1 \times
5^{\prime}.1$ FOV. All three telescopes are equipped with a 1024 $\times$
1024 pixel CCD. Table 1 displays the log of the performed observations for
GRB 980329.  The images were debiased and flat-fielded in the standard way
using the IRAF reduction package.  Magnitudes for the OT and for other
objects in the field were obtained using the SExtractor software based on
the corrected isophotal magnitude, which is very useful for star-like
objects (Bertin and Arnouts 1996).  In order to compare the magnitude of
the object with others R-band measurements available on 29-30 March, we
made use of the local standard considered by Palazzi et al. (1998), who on
29.99 March reported R = 23.6 $\pm$ 0.2 mag for the OT.  This star is
located at $\alpha_{2000} = 7^{h} 2^{m} 39.40^{s}$, $\delta_{2000} =
38^{\circ} 50^{\prime} 3.1^{\prime \prime}$, and has a magnitude R = 18.29
$\pm$ 0.05 mag (Palazzi 1998).  It has to be pointed out that this
calibration is systematically shifted ~$0.15$ mag with respect to the one
reported by Rhoads et al. (1998).

\begin{table}[t]
\begin{center}
\caption[Log of observations covering the GRB 980329 error box]
{\label{980329log} Log of optical observations covering the GRB 980329
  error box}
\begin{tabular}{cccll}
\hline
{\small Date of 1998}&{\small Filter}&{\small Exposure
  }&{\small Telescope} &{\small Limiting}\\
{\small (starting time)    }&{\small       }&{\small time (s)}&{\small
  }&{\small mag. ($3\sigma$)}\\ 
\hline
{\small 29.85 Mar}&{\small R}&{\small  900}&{\small 1.0-m OGS}&{\small R=22.2}\\
{\small 29.89 Mar}&{\small R}&{\small 2700}&{\small 1.0-m JKT}&{\small R=22.8}\\
{\small 30.88 Mar}&{\small R}&{\small 2700}&{\small 1.0-m JKT}&{\small R=20.3$^{\star}$}\\
{\small 31.00 Mar}&{\small R}&{\small 1800}&{\small 1.5-m OAN}&{\small R=20.8$^{\star}$}\\
{\small 31.83 Mar}&{\small R}&{\small 3600}&{\small 1.5-m OAN}&{\small R=21.2$^{\star}$}\\
{\small 31.93 Mar}&{\small R}&{\small 1800}&{\small 1.0-m JKT}&{\small R=22.8}\\
{\small 31.96 Mar}&{\small R}&{\small 1800}&{\small 1.0-m JKT}&{\small R=22.8}\\
{\small  2.89 Apr}&{\small R}&{\small 7200}&{\small 1.5-m OAN}&{\small R=21.5$^{\star}$}\\
{\small  2.98 Apr}&{\small B}&{\small 2400}&{\small 1.5-m OAN}&{\small B=20.5$^{\star}$}\\
\hline
\multicolumn{5}{l}{$\star$ = Observations carried out in bad weather conditions (clouds).}\\
\hline
\end{tabular}
\end{center}
\end{table}

The images taken with both the 1.0-m  OGS and 1.5-m  OAN telescopes did not
reveal the counterpart.  However, the JKT image  taken on 29.89 March shows
the presence of a faint  object located at the  position  of the radio  and
optical GRB counterparts (Taylor et al.  1998b, Klose  1998).  See Fig.  1. 
Astrometry of the object was carried out by means of  15 stars in the image
whose  coordinates were taken from  the USNO-A2.0 catalogue. Thus we obtain
$\alpha_{2000} = 7^{h}  2^{m}   38.02^{s} \pm 0.05^{s}$,  $\delta_{2000}  =
38^{\circ} 50^{\prime} 44.2^{\prime \prime} \pm 0.6^{\prime\prime}$ for the
counterpart, which is $<$ $0\farcs  5$ away from  the position of the radio
counterpart. The magnitude of the  OT at this time  was R = 23.7 $\pm$  0.5
mag.

Our detection on 29.89  March is the  second earliest optical detection  of
GRB 980329, corresponding the earliest one to the I-band detection by Klose
(1998).  As  it can be  seen on Fig.  2, this value  is consistent with the
measurements  reported by other  authors, namely R =  23.6 $\pm$ 0.2 mag on
29.99 March,  R = 25.2 $\pm$ 0.3  mag on 1.125  April (Palazzi et al. 1998)
and R = 25.7 $\pm$ 0.3 mag on 2.0 April (Djorgovski et al. 1998).

\section{Discussion}

\subsection{Reliability of the object}
A quantitative evaluation of the confidence level of our 2$\sigma$
detection by means of SExtractor is done following two methods.  First, we
considered the peak to peak signal-to-noise ratio, resulting that the peak
intensity of the object is 4.5$\sigma$ above the background level. On the
other hand, aperture photometry measurements for the object and for several
random points in the surrounding area were obtained by means of a FORTRAN
code developed by us. Then, we calculated the confidence level in the same
way as we did for the first case but using the integrated fluxes instead of
the peak intensities.  When aperture photometry of the object was carried
out considering radii ranging from 3 to 6 pixels, the confidence level of
the detection lies in the 3.4-2.0$\sigma$ range. As expected, enhancing the
aperture makes the background to come in, impoverishing the signal-to-noise
ratio. The peak signal-to-noise ratio only takes into account the maximum
value of the photodensity profile, so usually it will be larger than the
one obtained with the aperture photometry.

The OT magnitude and the 3$\sigma$ limiting magnitudes displayed in Table 1
have been calculated considering the most unfavourable case of a 6 pixels
radius, where the detection of the object is at a 2.0$\sigma$ level. This
fact explains why the magnitude of the object is below the quoted 3$\sigma$
limiting magnitude of R = 22.8 mag.  If a gaussian fit is made to the
object photodensity profile, we obtain a full width at half maximum (FWHM)
of 3 pixels (or 1$^{\prime \prime}$), which is the typical value of the
seeing derived for other objects in the field.  On the basis of the
confidence levels obtained, we consider the detection as {\em real.}

\begin{figure}[t]
 \centering
 \resizebox{\hsize}{!}{\includegraphics[angle=-90]{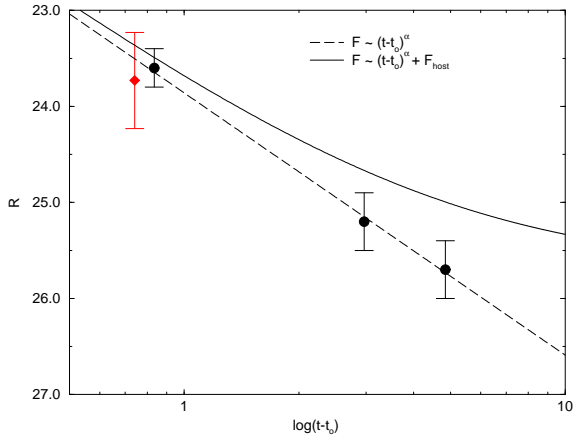}}
\caption{\label{980329fig4} The R-band light curve of the GRB 980329
  counterpart.  The diamond represents our detection on 29.89 March (R =
  23.7 $\pm$ 0.5 mag), and the circles are the values reported elsewhere,
  namely R = 23.6 $\pm$ 0.2 mag on 29.99 March, R = 25.2 $\pm$ 0.3 mag on
  1.125 April (Palazzi et al.  1998b), and R = 25.7 $\pm$ 0.3 mag on 2.0
  April (Djorgovski et al. 1998c).  The observed flux $F$ is very well
  fitted by a power law in the form; $F \sim (t-t_{o})^{\alpha}$ (long
  dashed line) with $\alpha = -1.1 \pm 0.4$.  If we consider the
  contribution of a host galaxy $F_{\rm host}$ with R = 25.7 mag, the fit is
  not so good. This suggests that {\em still} on 2.0 April the OT
  contributes to a large fraction of the optical flux, while the
  contribution of the host galaxy is negligible.}
\end{figure}

\subsection{Estimation of the magnitude of the host galaxy} 

By fitting a power law to the observed  flux $F \sim (t-t_{o})^{\alpha}$ we
obtain  an  index   $\alpha  =  -1.1  \pm    0.4$.    The  value   of   the
reduced-chi-squared $\chi^2/dof$ for the fit is  $\chi^2/dof = 0.15$, where
$dof$ is  the number of  degrees of freedom. We note  that the low value of
the reduced-chi-squared is due to   the fact that   the errors in the  data
points have been  overestimated.  This value is  in agreement  with the one
given by Palazzi et al. (1998), who obtained $\alpha = -1.3 \pm 0.2$.


As it can be seen in Fig. 2, the expected light curve for a R = 25.7 mag
host galaxy (Djorgovski et al. 1998) does not match well the observed flux.
This suggests that the total optical flux was still dominated by the
contribution of the OT when its magnitude was R = 25.7 mag on 2.0 April. In
fact, if we extrapolate a single power law decay from the first three
points to the one of 2.0 April, a value of R = 25.8 mag is obtained for the
OT.  This value is consistent, within the uncertainties, with the value
measured at that date (R = 25.7 $\pm$ 0.3 mag).



If we consider now the contribution of the host galaxy to the observed flux
$F=F_{\rm host} + B (t-t_{o})^{\alpha}$, we can estimate the magnitude of the
host galaxy.  The values of $\chi^{2}/dof$ exhibits a broad minimum around
R = 27.4 mag and so we conclude that the magnitude of the GRB 980329 host
galaxy should lie in the range given by 26.8 mag $<$ R $<$ 29 mag.



\subsection{Similarities between GRB 980329 and GRB 970111}

GRB 970111 and GRB 980329 are two of the most intense GRBs detected by {\it
  BSAX}, showing prominent emission above 40 keV.  In fact, their fluence
in the 50-300~keV range is more than four times larger than the other GRBs
detected (with the exception of GRB 990123).  On the other hand, they
displayed the hardest spectra, showing a hardness ratio (HR) between
0.5$\pm$0.2 and 1.0$\pm$0.2 (in\' \rm t Zand et al. 1998). Therefore, at
first sight one could speculate that both GRBs were originated under
similar physical conditions and nearer than the other GRBs of the {\it
  BSAX} sample.

If this was the  case, the optical decay  curves should be somehow similar. 
This fact could give us a clue for  explaining the non detectability of GRB
970111 OT  $\sim$19  hr after the  gamma-ray  event (Castro--Tirado et  al. 
1997).   Assuming a  similar decay trend   and  apparent magnitude for  GRB
970111 than those of  GRB 980329, the magnitude of  the GRB 970111 OT 19 hr
after the  high energy event would  be  R $\sim$ 23.5  mag,  just below the
detection limit  of our images  taken  on 20 Jan,   1997 (Gorosabel et al.  
1998).

It could be possible that  both GRBs were  originated from a host galaxy at
$z < 1$, a nearby  source in comparison  to the other  {\it BSAX} GRBs  for
which redshifts  has been measured.  In fact,  Palazzi et al.   (1998) have
suggested   that GRB  980329  could be    arised  from a strongly  obscured
starburst galaxy at $z \sim 1$.

This statement could be supported if a correlation between the $\gamma$-ray
flux and distance  of GRBs is present. However,  such a correlation has not
been  detected    among  the BSAX   GRBs  sample.   Furthermore, it  is  in
contradiction to the   photometric redshift estimation  of Fruchter (1999),
who proposes that GRB 980329 host galaxy is at $z \sim 5$.

\section{Conclusion}

We present the optical follow up observations performed for GRB 980329 by
means of the data acquired at the OGS, OAN and JKT telescopes.  Only the
image taken on 29.89 March with the JKT telescope reveals an object within
$0\farcs 5$ from the OT and radio positions.  The magnitude of the OT on
29.89 March was R = 23.7 $\pm$ 0.5 mag, which is consistent with R = 23.5
$\pm$ 0.2 mag on 29.99 March, reported by Palazzi et al. (1998).

Based on extrapolations of the light curve we claim that the R-band
magnitude of the GRB 980329 host galaxy should lie in the range 26.8 mag
$<$ R $<$ 29 mag, considerably fainter as claimed by Djorgovski et al.
1998. This prediction should be tested with deep ground-based or {\it HST}
observations.

GRB 980329 shows similar characteristics to GRB 970111.  Assuming a similar
decay trend and apparent magnitude, the  OT associated to  GRB 970111 19 hr
after  the high-energy  event would  be R $\sim$   23.5 mag, just below the
detection limit of our data.  This fact could  explain the non-detection of
the GRB 970111 OT.

\begin{acknowledgements}
  We whish to thank the referee, J. Greiner for his valuable comments and
  suggestions. We are also indebted to M. Tafalla for his valuable support
  at the 1.5-m OAN telescope.  We are very grateful to Eliana Palazzi and
  Nicola Masseti for valuable information on the calibration stars. This
  work has been partially supported by Spanish CICYT grant
  ESP95-0389-C02-02.

\end{acknowledgements}

\bibliographystyle{aa}

\end{document}